# Photonic light trapping in silicon nanowire arrays: deriving and overcoming the physical limitations

Sebastian W. Schmitt[1,2*] and Silke H. Christiansen[1,2,3]

[1] Helmholtz-Zentrum Berlin für Materialien und Energie, Institute Nanoarchitectures for Energy Conversion, Hahn-Meitner-Platz 1, 14109 Berlin / Germany

[2] Max Planck Institute for the Science of Light, Photonic Nanostructures, Günther-Scharowsky-Str. 1, 91058 Erlangen / Germany

[3] Freie Universität Berlin, Department of Physics, Kaiserswerther Str. 16-18, 14195 Berlin / Germany

*Corresponding author: sebastian.schmitt@mpl.mpg.de, sebastian.schmitt@helmholtz-berlin.de

**Abstract:** Hexagonally aligned, free-standing silicon nanowire (SiNW) arrays serve as photonic resonators which, as compared to a silicon (Si) thin film, do not only absorb more visible (VIS) and near-infrared (NIR) light, but also show an inherent photonic light concentration that enhances their performance as solar absorbers. Using numerical simulations we show, how light concentration is induced by high optical cross sections of the individual SiNWs but cannot be optimized independently of the SiNW array absorption. While an ideal spatial density exists, for which the SiNW array absorption for VIS and NIR wavelengths reaches a maximum, the spatial correlation of SiNWs in an array suppresses the formation of optical Mie modes responsible for light concentration. We show that different from SiNWs with straight sidewalls, arrays of inverted silicon nanocones (SiNCs) permit to avoid the mode suppression. In fact they give rise to an altered set of photonic modes which is induced by the spatial correlation of SiNCs in the array, and therefore show a higher degree of freedom to independently optimize light absorption and light concentration. Apart from explaining the good light absorbing and concentrating properties of SiNC arrays, the work justifies a revaluation of SiNW arrays as optical absorbers.

**Introduction:** To overcome the problem of low light absorption in silicon (Si) thin films, anti-reflective and light trapping surface structures were implemented in advanced photovoltaic (PV) device concepts, and modern nanotechnology could further promote innovation in the development of elaborate light management strategies. Random scatterers, like pyramidal structures[1], upright nanocones[2], periodic[3], quasi random[4,5], or randomly textured structures[6] have been investigated and could reduce reflection and enhance light trapping of Si thin films towards the ray optics limit (Lambertian limit / Yablonovich limit)[7]. In wavelength-scale photonic structures, like arrays of SiNWs[8], photonic crystals[9] or Si nanospheres[10] classical ray optics is no longer valid and they offer a fundamentally different approach towards light trapping strategies. Here, light trapping in photonic modes can even exceed the Lambertian limit for certain resonant wavelengths[11]. Photonic modes enhance the local density of states (LDOS) in the material, which permits to achieve higher PV device efficiencies caused by light concentration[11–14]. Arrays of SiNWs were proposed to improve light absorption in Si in both different ways: while random multiple scattering in SiNW arrays enhances broadband light absorption[15–17], resonant optical phenomena such as Mie scattering at individual SiNWs[18–20] constitute the optical response of the entire absorber, i.e. cause an overall wavelength-, angle- or polarization- selective photonic absorption enhancement[21–23]. Recently, arrays of inverted SiNCs were demonstrated to even exceed the light trapping in SiNW arrays[24].

Using the results of numerical simulations, this work presents strategies to optimize light absorption in free standing arrays of SiNWs and SiNCs and specifies the fundamental difference between the two geometries. An explicit focus resides on the limitations of inherent light concentration (photonic light trapping) in arrays of SiNWs and SiNCs. It is demonstrated that once arranged in arrays, the SiNCs show a higher VIS-NIR absorption while simultaneously inducing a much higher intrinsic light concentration than SiNWs, which are less favorable but were already widely discussed in this context[8,18,25–30]. The important difference in absorption efficiency can be attributed to the photonic mode formation within SiNC arrays which is completely different from the mode formation SiNW arrays. While SiNC arrays promote photonic resonances in neighboring SiNCs, resonances in neighboring SiNWs extinct one another and do not permit a simultaneous optimization of light absorption and concentration. The SiNC array delineates therefore a prototype of a photonic resonator structure which is suitable for the design of controllable, inherently light concentrating, Si-based light absorbers and as such can serve as a very efficient building block in thin-film solar cell or solar-fuel device concepts. The first section of this work serves as an introduction to all necessary theory, required to appreciate the presented study. On this basis, the main claims of the manuscript will carefully be derived.

**Light absorption and concentration in individual silicon nanowires:** An individual SiNW with a diameter in the range of VIS to NIR light can be regarded as a photonic antenna and therefore exhibits a highly nonlinear absorption and scattering behavior[21]. Its light absorption / scattering properties can be described by the so-called absorption / scattering cross section $\sigma_{abs} / \sigma_{sca}$ which is generally defined as[31]

$$\sigma_{abs/sca} = \frac{P_{\gamma,abs/sca}}{j_\gamma} \qquad (1)$$

where $j_\gamma$ is the incident photon current density and $P_{\gamma,abs/sca}$ denotes the total power absorbed / scattered by the SiNW. Since $j_\gamma$ is given in $W/m^2$ and the total absorbed / scattered power has the unit $W$, $\sigma_{abs} / \sigma_{sca}$ intuitively represents the area, in which the the SiNW absorbs / scatters the incoming light. $\sigma_{abs} / \sigma_{sca}$ is dependent on wavelength, polarization, angle of incidence, diameter and length of the SiNW since these parameters are also characteristic for the photonic modes that can be hosted by the SiNW optical antenna. $\sigma_{abs} / \sigma_{sca}$ can significantly be higher than the geometrical projection area $A_p$ of the SiNW in the plane perpendicular to the direction of the incident light. To describe the light concentration on $A_p$, which occurs along with light absorption of the SiNW, a concentration factor $X$ can be calculated according to

$$X = \frac{\sigma_{abs}}{A_p} \qquad (2)$$

Physically, this concentration corresponds to an enhancement of the LDOS in the material. Figure 1a shows a numerical simulation of the spectral absorption cross section $\sigma_{abs}$ for free floating individual SiNWs with a diameter of 350nm and lengths of 1.6, 2.4, 3.2, 6.4 and 9.6µm illuminated by a linearly polarized plane wave (300-1100nm) incident along the SiNW axis. Here, $\lambda_o = 300$nm roughly marks the onset of the solar spectrum and $\lambda_1 = 1100$nm (1.12eV) corresponds to the band gap of Si, above which no absorption will occur. Integration of the spectra shows that the average absorption cross section

$$\langle \sigma_{abs} \rangle = \frac{1}{\lambda_1 - \lambda_o} \int_{\lambda_o}^{\lambda_1} \sigma_{abs}(\lambda) d\lambda \qquad (3)$$

rises with an increasing length of the SiNWs ($l = 1.6, 2.4, 3.2, 6.4, 9.6$µm, $\langle \sigma_{abs} \rangle = 0.19, 0.25, 0.30, 0.49, 0.66$µm$^2$). Further, the spectrum of the longest SiNW is dominated by four pronounced spectral peaks, which decrease in intensity and show a slight blue shift as the SiNW gets shorter. Figure 1c shows $\sigma_{abs}$ for SiNWs with a length of 1.6µm and diameters $d$ of 100, 180, 350, 540 and 760nm for the same irradiation conditions. Here, $\langle \sigma_{abs} \rangle$ rises with with increasing diameter ($d = 100, 180, 350, 540, 760$nm, $\langle \sigma_{abs} \rangle = 0.05, 0.10, 0.19, 0.30, 0.47$µm$^2$) and so does the magnitude of the most pronounced spectral peaks. Comparing Figures 1a and c, the height, number and spectral position of those peaks is obviously characteristic for the SiNW diameter i.e. they correspond to so-called Mie modes propagating in a plane perpendicular to the SiNW axes. Spectra in Figure 1c indicate that a higher number of (less confined) modes can be hosted by SiNWs with a larger diameter. The example of a SiNW with 350nm diameter and 1.6µm length in Figure 1a indicates that $\sigma_{abs}$ and $\sigma_{sca}$ usually peak at about the same spectral positions i.e. where the SiNW interacts most efficiently with the incoming radiation. This fact is not immediately relevant for the absorption of individual SiNWs but helps to explain the absorption properties of SiNW arrays, as will be discussed below.

Figures 1b, d illustrate the spectral light concentration $X$ for the free floating individual SiNWs from Figures 1a, c calculated according to Equation 2. Integration of the spectra in Figure 1b shows that the average light concentration between $\lambda_o = 300$nm and $\lambda_1 = 1100$nm

$$\langle X \rangle = \frac{1}{\lambda_1 - \lambda_o} \int_{\lambda_o}^{\lambda_1} X(\lambda) d\lambda \qquad (4)$$

rises with an increasing length of the SiNW ($l = 1.6, 2.4, 3.2, 6.4, 9.6$µm, $\langle X \rangle = 1.9, 2.6, 3.1, 5.1, 6.9$) which is intuitively clear since all SiNWs have the same $A_p$ while, as stated above, $\langle \sigma_{abs} \rangle$ is increasing with length. However, with increasing diameter the trend for $\langle X \rangle$ as compared to $\langle \sigma_{abs} \rangle$ is opposite. Integration of spectra in Figure 1d according to Equation (4) results in a decreasing $\langle X \rangle$ with increasing SiNW diameter ($d = 100, 180, 350, 540, 760$nm, $\langle X \rangle = 6.3, 3.7, 1.9, 1.3, 1.0$). *Therefore it can be concluded that for the regarded range of wavelengths and geometries, individual SiNWs absorb more light relative to their projected area $A_p$, i.e. effectively demonstrate light concentration as defined in Equation 2 that is increasing with increasing aspect ratio ($AR = l/d$).* Note that a higher light concentration only corresponds to a higher absolute light absorption (i.e. $\langle \sigma_{abs} \rangle$) in the SiNW, once SiNWs of the same $A_p$ (or $d$) with different $l$ are compared.

Another feature of the spectra shown in Figure 1 deserves further attention. With decreasing aspect ratio $AR$ of the SiNWs, spectra for $\sigma_{abs}$ and $X$ exhibit an increasing number of low bandwidth peaks above 600nm. These peaks originate from Fabry-Perot (FP) oscillations along the SiNW axis which rise in relative intensity, once the SiNW $AR$ decreases. This can be understood looking at Figure 2. Here, the absorption depth $\delta$ of light in Si is plotted with respect to the wavelength $\lambda$, while the same graph is used to illustrate the SiNW length $l$ over diameter $d$. If the wavelengths are longer than the SiNW diameter ($\lambda > d$), or if the wavelengths absorption depth is shorter than the length of the SiNW ($d(\lambda) < l$), absorption enhancement / concentration due to FP modes guided along the SiNW axis will be minor. For $\lambda > d$, only a limited number of FP modes will fit the SiNW core and therefore these modes do not extensively contribute to absorption. On the contrary, for $\delta(\lambda) < l$, FP modes will significantly be damped for light propagation along the SiNW axis with length $l$. Practically this means that FP modes will only show a considerable contribution to light absorption in SiNWs for all geometries below the red line in Figure 2. Otherwise the absorption enhancement will be dominated by Mie modes. The inset in Figure 2 depicts the mode formation in a SiNW with $d = 760$nm and $l = 1.6$µm for a 455nm and a 793nm incident plane wave. While the short wavelength with a low penetration depth can only form a (Mie) mode perpendicular to the SiNW axes, the longer wavelength forms a FP mode propagating throughout the entire SiNW. The spectral response of the two modes is indicated with red stars in the uppermost $\sigma_{abs}$ spectrum of Figure 1c. As previously stated, the Mie mode appears as one of the characteristic peaks in the short wavelength regime while the FP mode produces a narrow peak at a higher wavelength. In Figure 2, the geometries of the SiNWs which are topic of Figure 1 are indicated. Apart from the SiNW with $d = 760$nm, and $l = 1.6$µm all these geometries fulfill the conditions for a low relative intensity of FP modes at wavelengths larger than 600nm i.e. are located above the red line in Figure 2. The high absorption cross sections / concentration factors (see e.g. Figure 1d, the SiNW with $d = 100$nm absorbs light of wavelength $\lambda = 586$nm on an area which is about 40 times larger than its geometrical projection area $A_p$), also give evidence of an enormous diffraction of light out of its incident direction into the plane perpendicular to the SiNW long axis[13]. The diffraction is caused by the Mie modes, which propagate in this plane and as mentioned before account for the bulk of the resonant absorption enhancement in most SiNW geometries considered in Figures 1 and 2.

**Light absorption and concentration in silicon nanowire arrays:** With only slight modifications the aforementioned definitions of $\sigma_{abs}$ and $X$ for individual SiNWs also apply to SiNW arrays. For a unit area $\hat{a}$ of the SiNW array, $\sigma_{abs}$ can be written as in (1)

$$\sigma_{abs} = \frac{P_{\gamma,abs}}{j_\gamma} = \frac{j_{\gamma,abs} \cdot \hat{a}}{j_\gamma} = A \cdot \hat{a} \qquad (5)$$

with $j_{\gamma,abs} / j_\gamma$ being the absorption $A$ of the SiNW array. The geometrical projected part $A_p$ of the SiNW array unit area $\hat{a}$ (perpendicular to the SiNW axes) can be found by multiplying it with the array filling fraction $f$. With

$$A_p = f \cdot \hat{a} \qquad (6)$$

$X$ for the SiNW can be derived as

$$X = \frac{\sigma_{abs}}{A_p} = \frac{A}{f} \qquad (7)$$

In an array of SiNWs, as can already be seen from equation (7), concentration and absorption occur in the same nanoscale object and both properties are directly interrelated. From (7) a fundamental design rule for a SiNW array absorber can already be derived: *The ideal geometry of a photonic SiNW array absorber exhibits the highest absorption $A$ at the lowest filling fraction $f$*. Furthermore, it can be noted: if a SiNW array of height $l$ has the same absorption $A$ as a Si slab of thickness $l$, it is the better absorber since it absorbs the light on an area reduced by the inverse filling fraction $1/f$ and therefore exhibits inherent light concentration which increases the voltage of the device (assuming a similar electrical performance of the device constituting Si diode)[13,14].

In the following, an optimization strategy of light absorption/concentration in hexagonally aligned SiNW arrays is derived. Therefore, a numerical analysis of the SiNW array absorption (reflection, transmission) is carried out. It is based on hexagonally aligned SiNW arrays which are variable in the SiNW diameter $d$, the length $l$, and the pitch $a$, and a linearly polarized plane wave (300-1100nm) incident along the SiNW axis. In the simulations, FP modes in the SiNWs will be suppressed by embedding their lower end into a perfectly matching layer (PML) provided by the simulation software (Lumerical, FDTD solutions). This simplification is valid, since according to Figure 2, FP modes play only a minor role for absorption enhancement in most of the considered geometries. The discussion starts with Figures 3a, b and c which shows the spectrally averaged, light absorption $\langle A \rangle$, transmission $\langle T \rangle$ and reflection $\langle R \rangle$ of a hexagonally aligned SiNW array with 1.6µm length and a pitch of 500nm as a function of SiNW diameter $d$. Here, and in the following, the spectrally averaged values of absorption $\langle A \rangle$, reflection $\langle R \rangle$ and transmission $\langle T \rangle$ of the SiNW arrays between $\lambda_o = 300$nm and $\lambda_1 = 1100$nm will be calculated as

$$\langle A, R, T \rangle = \frac{1}{\lambda_1 - \lambda_o} \int_{\lambda_o}^{\lambda_1} A, R, T(\lambda) d\lambda \qquad (8)$$

It can be observed that $\langle A \rangle$ peaks for a specific diameter $d_{max}$, at a maximum value which is significantly higher than $\langle A \rangle$ for a Si slab of 1.6µm thickness. Note that $\langle A \rangle$ of the slab is calculated for a single pass of incident light, disregarding FP modes as for the SiNW arrays. The grey dashed lines in Figures 3b and c show a linear estimate of $\langle T \rangle$ and $\langle R \rangle$ of the SiNW arrays using the values for the Si slab multiplied by the filling fraction $f$ of the SiNW array with 500nm pitch and the associated SiNW diameter. The comparison of this linear estimate with the numerically calculated values of $\langle T \rangle$ and $\langle R \rangle$ shows that $\langle T \rangle$ of the SiNW array is much lower than could be explained by the filling fraction, while $\langle R \rangle$ roughly resembles its linearly estimated value. Accordingly, as can also be seen by a comparison of the exemplary spectra for $A$, $R$ and $T$ in Figures 3d, e and f ($d = 150, 350, 480$nm), the transmission (and therefore the absorption $A = 1 - R - T$) of the SiNW arrays exhibits a much stronger nonlinear behavior than the reflection.

On the one hand, the nonlinear absorption of the SiNW arrays can be attributed to the strongly nonlinear absorption cross section $\sigma_{abs}$ of individual SiNWs. Note, however that for the SiNWs arranged in arrays, also the scattering cross section $\sigma_{sca}$ implicitly contributes to the mode-induced absorption enhancement. Even though the back scattering / reflection from the arrays is not strongly influenced by the modes formed in the SiNWs (compare Figure 3f), the forward and sideward scattering into the SiNW array follows $\sigma_{sca}$, which peaks at about the same wavelengths as $\sigma_{abs}$ (Figure 1a, grey line / right axis). In this case it is reasonable to define an extinction cross section

$$\sigma_{ext} = \sigma_{abs} + \sigma_{sca} \qquad (9)$$

which describes the total interaction of individual SiNWs with the incident light. For large SiNW diameters ($d \to a$) the values of $\langle A \rangle$, $\langle T \rangle$ and $\langle R \rangle$ as well as the corresponding spectra in Figures 3d, e, f converge to the results for the Si thin film as expected. The maximum of $\langle A \rangle$ occurs, as soon as the ideal ratio between $\langle R \rangle$ and $\langle T \rangle$ is reached.

We continue the discussion with the effect of the array pitch and the SiNW diameter on the absorption properties of SiNW arrays. Figure 4a shows the spectrally averaged, light absorption $\langle A \rangle$ of hexagonally aligned SiNW arrays with 1.6µm length and pitches of 125, 250, 500, 750, and 1000nm as a function of SiNW diameter $d$. It can be seen that for the variation of array pitches, $\langle A \rangle$ peaks for a specific diameter $d_{max}$. While for $a$ between 250 and 1000nm the maximum value for $\langle A \rangle$ is about 0.1 higher than the value of the corresponding thin film, the value for $a = 125$nm is considerably lower which will be explained in detail later in the text. The specific diameter $d_{max}$ for which $\langle A \rangle$ reaches the maximum stays constant as SiNWs in the array get longer. Figure 4b displays the example for $a = 500$nm and an increasing length from 1.6 to 9.6 µm. It can be observed that the value of $d_{max}$ is 350nm, independently of the length of the SiNWs which constitute the array. However, the peak in $\langle A \rangle$ over the diameter $d$ gets slightly broader which can be explained by the fact that in longer arrays also the light scattered by the SiNWs will increasingly be absorbed. Plotting the SiNW diameter at the maximum $\langle A \rangle$ of a SiNW array with length 1.6µm ($d_{max}$) as a function of pitch $a$ a linear behavior can be derived (Figure 4 c). A linear fit of the data passing through the origin results in the relation

$$d_{max}(a) = 0.74 \cdot a \qquad (10)$$

Accordingly, we can calculate the filling fraction of a SiNW array with the highest averaged light absorption $\langle A \rangle$ between 300nm and 1100nm as being

$$f = \frac{A_p(d_{max}(a))}{A_{hex}(a)} \approx 0.5 \qquad (11)$$

Here, $A_p(d_{max}(a))$ is the projection area of the SiNW with $d_{max}(a) = 0.74 \cdot a$ and $A_{hex}(a)$ is the area of the hexagonal unit cell of the array with pitch $a$. Due to the linear relation between $d_{max}$ and $a$ this value is constant, which leads to further fundamental design rules for SiNW absorbers. *For homogenous illumination with wavelengths between 300 and 1100nm, the maximum absorption in a SiNW array of finite length is reached for a filling fraction of $f \approx 0.5$. According to Equation 7 it can be concluded that the maximum spectral light concentration $X$ for the ideally absorbing SiNW array cannot exceed a value of about 2, since the value of A for an arbitrary wavelength cannot be higher than 1.*

Nevertheless, the spectral light concentration in non-ideally absorbing SiNW arrays can go beyond a value of 2. This is further illustrated by Figure 5. In panels a, b (c, d) the spectral light absorption $A$ and the spectral light concentration $X$ of hexagonally aligned SiNW arrays with 1.6µm length, 350nm diameter and a pitch of 500, 750 and 1000nm (100nm diameter and a pitch of 125, 250 and 500nm) is shown. Once the pitch exceeds a value of 500nm (125nm), which according to the previous considerations corresponds to the ideally absorbing configuration, the light absorption in the SiNW array decreases and the spectral light concentration as calculated by Equation 7 grows. For a pitch as large as 1000nm (500nm) the concentration in the array actually approaches the spectral light concentration of an individual SiNW. Figures 5a (5c) further give evidence that the nonlinear optical behavior of a SiNW array is directly induced by the nonlinear extinction cross section of its individual constituents. For a high pitch within the array, the peaks in the extinction cross section of individual SiNWs directly match the peaks of the array absorption. Once the array pitch decreases, these peaks get less pronounced, and additionally show an overall blue shift. Figures 5e and f demonstrate the spectral light absorption $A$ and the spectral light concentration $X$ of hexagonally aligned SiNW arrays with 500nm length, 350nm diameter and a length of 1.6, 3.2 and 6.4µm. Intuitively, the absorption of the array increases with increasing length of the SiNWs, while as previously stated the spectral concentration approaches $f \approx 2$ for the longest array. The spectral concentration of the individual SiNWs constituting the longest array is plotted for comparison. As expected, it also by far exceeds the value for the array which again was chosen according to the design rules for the highest absorption ($a = 500$nm, $d = 350$nm).

From the findings it can be concluded that SiNWs, once embedded in an array configuration, loose great parts of their light concentration capacity due to the vicinity to their neighboring structures that cause destructive interference. Figures 6a and b show the relative cross sectional energy ($E^2$) density of adjacent SiNW for the array configurations shown in Figures 5a, b and 5c, d at the wavelengths that correspond to the highest (for smaller pitches blue-shifting) peak between 500 and 600nm wavelength. It can clearly be observed that for a smaller pitch the transversal Mie modes are more and more limited to the top part of the SiNW arrays. For the lowest pitch in Figure 6b ($a = 125$nm) the formation of Mie modes is hardly possible since the incoming light cannot penetrate into the 'voids', i.e. the free area between SiNWs, of the array at all. This finding explains, why the light concentration in the SiNW array decreases with an increasing filling fraction (decreasing pitch).

*Due to the limited penetration of light into the voids in dense arrays of SiNWs, the contribution of Mie modes to absorption tends to be weaker with increasing filling fraction i.e. the peaks in the absorption and concentration spectra smear out (see Figures 5a-d).* Note that this effect also explains the low peak in the spectrally averaged absorption of the $a = 125$nm array in Figure 4a, and that there is no sense in discussing photonic light absorption enhancement in SiNW arrays of finite length with a pitch a below this value. Further, the mode suppression in SiNW arrays explains the linear dependence of $d_{max}$ on $a$ (Figure 4c, Equation 10). Since at the diameter of maximum absorption the SiNW arrays for any pitch are already quite dense, the absorption maxima in the spectra which appear due to photonic resonances are already smeared out and account for less than 10 percent of the total absorption in the array (compare Figure 3d, green line). In contrast, for the absorption of an array with lower density (e.g. red line Figure 3d in the manuscript, $d = 150$nm, l=1.6µm, a= 500nm), the photonic resonances make up the major part of the overall absorption. This leads to the conclusion that for the maximum absorption of the SiNW arrays, the contribution of nonlinear effects (which depend on the geometry of the SiNW in the array) is minor, and the resulting absorption is merely caused by an optimum between top facet reflection and quasi-bulk absorption in the array.

The previous considerations present fundamental design rules for tailoring light absorption and concentration in individual SiNWs and SiNW arrays. While an optimization strategy for the absorption enhancement could be found, it counteracts the ideal requirements for light concentration that is caused by photonic resonances. While in individual SiNWs transversal Mie modes are responsible for the strongest photonic effects, modifications of the same modes could explain the nonlinear optical properties of SiNW arrays. Once SiNWs are arranged in dense arrays, it could be observed that optical interaction of adjacent SiNWs restrict the photonic mode formation and confines it to the very top of SiNWs. This causes deviations and finally a suppression of the mode profile of individual, array-constituting SiNWs.

These findings raise the question if it is possible to design arrays of photonic Si absorbers that show an inherent mode profile, which is independent of the mode formation in the individual array-constituting absorbers. The following section will give an example of an array of photonic Si absorbers that fulfills this requirement and is therefore of utmost importance for any consideration of one-dimensional structures constituting solar absorbers e.g. in thin-film solar cell or solar fuel device concepts.

**Photonic light trapping in arrays of silicon nanocones:** In this section it is demonstrated that an array of SiNWs which is modified to an array of inverted silicon nanocones (SiNC), i.e. one-dimensional Si nanostructures with a slight, negative sidewall taper, shows an enhanced absorption, even at a lower array filling fraction. Furthermore, the SiNC arrays exhibit an inherent mode profile, which is independent of the modes which prevail in an individual inverted SiNC. Figure 7 shows the essence of our systematic investigation for the consideration of one-dimensional Si nanostructures as solar light absorbers. It displays the spectrally averaged light absorption $\langle A \rangle$, transmission $\langle T \rangle$ and reflection $\langle R \rangle$ of SiNWs / SiNCs with 350nm radius and 1.6µm length arranged in a hexagonal lattice with a pitch of 500nm as a function of SiNWs / SiNCs bottom diameter. The initial SiNW array geometry was chosen to account for optimal absorption between 300 and 1100nm as discussed in Figure 4. In the graph, the bottom diameter $d$ is plotted in units of the top diameter $D$. As $d$ decreases to about 50% of the top diameter, the array absorption continuously grows towards an optimum value. This optimum value is reached, once an increasing light reflection compensates for the enhanced light absorption (i.e. lower transmission, $A = 1 - R - T$) with reducing bottom diameter.

The reasons for the higher spectrally averaged absorption of the inverted SiNC array as compared to the SiNW array is described in more detail by Figure 8. It shows the spectral light absorption $A$ of SiNWs with 350nm diameter and 1.6µm length arranged in a hexagonal lattice with a pitch of 1000nm (black) and $A$ of the same array with a reduced SiNW bottom diameter $d/D = 0.4$ (red). It can be observed that the improved light absorption of the SiNC array as compared to the SiNW array is expressed by a multitude of very sharp peaks in the absorption spectrum. As can be shown by the color-coded relative cross sectional energy density in adjacent SiNCs for three selected wavelengths ($\lambda = 560$, 757, and 945nm), the peaks are caused by the formation of strong photonic modes. Unlike in the array of SiNWs shown in Figure 6, these modes appear in different heights of the SiNC irrespective of a relatively high array density. According to Equation 7, the photonic absorption enhancement in the SiNC arrays is accompanied by an increased spectral light concentration, as shown in the right axis of Figure 8. Note that the filling fraction for the SiNC arrays is calculated, dividing the geometrical projection area $A_p$ (i.e. the area of the top facet) of an individual SiNC, by the hexagonal unit area $\hat{a}$ of the corresponding array. Therefore the array filling fraction for SiNC as well as for the SiNW arrays represents an area filling fraction. As it is obvious from the tapered shape of the SiNCs, the volumetric filling fraction of a SiNC array is even lower than for a SiNW array with the same top diameter and length. As per Equation 7, the light concentration for any wavelength in a SiNW / SiNC array cannot exceed $X = 1/f$ since the maximum absorption $A$ is 1. Accordingly, for the highest photonic light concentration the maximum absorption of a SiNW / SiNC array has to be reached at the lowest possible filling fraction.

Figure 9a shows the spectral light absorption $A$ of SiNCs with 350nm top diameter, bottom diameter $d/D = 0.4$ and 1.6µm length arranged in a hexagonal lattice with a pitch of 500 and 600nm. Surprisingly, unlike in an array of SiNWs, the absorption of the array with the higher pitch (lower filling fraction) is not substantially lower. This has the aforementioned consequences for the light concentration in the two SiNC arrays i.e. the one with the lower filling fraction shows evidence of a significantly higher light concentration (Figure 9b). The grey line shows $X$ of an individual SiNC with 350nm top diameter, bottom diameter $d/D = 0.4$ and 1.6µm length. The distribution of spectral peaks is completely different as compared to the SiNC arranged in an array. Furthermore, the individual SiNC shows a less pronounced light concentration in the NIR infrared range, which clearly demonstrates that some modes are an inherent consequence of the SiNC array configuration. The SiNCs therefore are an example of the aforementioned case, where the vicinity of individual array constituting structures is required for the efficient mode formation in the array. It proves a fundamentally new concept towards efficient light absorption in arrays of photonic absorbers, in which the periodic arrangement is key to the intended mode formation. Arrays of SiNCs and related structures with comparable photonic properties therefore require an in-depth understanding, so that the potential of inherent light concentration in arrays of Si nanostructures can fully be exploited.

**Conclusions:** The work investigates nanostructured solar light absorbers which act as inherently light concentrating structures (photonic light trapping). However, as the comprehensive numerical analysis shows, arrays of SiNWs are rather inappropriate for the design of light absorbers since they do not permit a simultaneous optimization of light absorption and concentration. This is most noteworthy, since there is a wealth of literature advertising SiNWs for enhanced solar light absorption. This literature usually infers optical properties of individual SiNWs to be applicable to arrays of SiNWs, a fact that was proven wrong by the present study. It is demonstrated that the spatial correlation of SiNWs in an array widely suppresses the formation of Mie modes which are responsible for the enhancement of light absorption in SiNW-based photonic absorbers[19,21]. The modification of SiNW arrays to arrays of inverted SiNCs permits to avoid this problem, giving rise to a totally altered set of photonic modes which is induced by the spatial correlation of SiNCs in the array. The SiNCs are a prototype of a photonic resonator structure which is suitable for the design of inherently concentrating absorbers.

**Acknowledgements:** The authors gratefully acknowledge financial support by the European Commission FP7: NMP priority - UNIVSEM (Nr. 280566), FIBLYS (Nr. 214042), and RODSOL (Nr. 227497); Health Priority - LCAOS (Nr. 258868) as well as by the German Research Foundation (DFG): FOR 1616. S.W.S. acknowledges T. Feichtner for assistance with Lumerical and R. Keding for taking care of the computing hardware.

**Contributions:** S.W.S conceived the study, performed the numerical simulations, evaluated the results and wrote the manuscript. S.W.S and S.H.C. discussed the manuscript at all stages.

**Additional information:** Competing financial interests: The authors declare no competing financial interests.

**Bibliography:**

1. Campbell, P. & Green, M. a. Light trapping properties of pyramidally textured surfaces. *J. Appl. Phys.* **62,** 243–249 (1987).
2. Jeong, S., McGehee, M. D. & Cui, Y. All-back-contact ultra-thin silicon nanocone solar cells with 13.7% power conversion efficiency. *Nat. Commun.* **4,** 2950 (2013).
3. Bozzola, A., Liscidini, M. & Andreani, L. C. Photonic light-trapping versus Lambertian limits in thin film silicon solar cells with 1D and 2D periodic patterns. *Opt. Express* **20,** 224–243 (2012).
4. Martins, E. R., Li, J., Liu, Y., Zhou, J. & Krauss, T. F. Engineering gratings for light trapping in photovoltaics: The supercell concept. *Phys. Rev. B* **86,** 041404 (2012).
5. Martins, E. R. *et al.* Deterministic quasi-random nanostructures for photon control. *Nat. Commun.* **4,** 2665 (2013).
6. Branz, H. M. *et al.* Nanostructured black silicon and the optical reflectance of graded-density surfaces. *Appl. Phys. Lett.* **94,** 1–4 (2009).
7. Yablonovitch, E. Statistical ray optics. *J. Opt. Soc. Am.* **72,** 899–907 (1982).
8. Kelzenberg, M. D. *et al.* Enhanced absorption and carrier collection in Si wire arrays for photovoltaic applications. *Nat. Mater.* **9,** 239–244 (2010).
9. Callahan, D. M., Munday, J. N. & Atwater, H. A. Solar cell light trapping beyond the ray optic limit. *Nano Lett.* **12,** 214–218 (2012).
10. Garín, M. *et al.* All-silicon spherical-Mie-resonator photodiode with spectral response in the infrared region. *Nat. Commun.* **5,** 3440 (2014).
11. Yu, Z., Raman, A. & Fan, S. Fundamental limit of nanophotonic light trapping in solar cells. *Proc. Natl. Acad. Sci. U. S. A.* **107,** 17491–17496 (2010).
12. Krogstrup, P. *et al.* Single-nanowire solar cells beyond the Shockley–Queisser limit. *Nat. Photonics* **7,** 1–5 (2013).
13. Xu, Y., Gong, T. & Munday, J. N. The generalized Shockley-Queisser limit for nanostructured solar cells. *Sci. Rep.* **5,** 13536 (2015).
14. Würfel, P. Intensity dependence of the efficiency in *Physics of Solar Cells : From Principles to New Concepts* 151 (Wiley-VCH, 2005).
15. Tsakalakos, L. *et al.* Silicon nanowire solar cells. *Appl. Phys. Lett.* **91,** 233117 (2007).
16. Garnett, E. C. & Yang, P. Light trapping in silicon nanowire solar cells. *Nano Lett.* **10,** 1082–1087 (2010).
17. Brönstrup, G. *et al.* Statistical model on the optical properties of silicon nanowire mats. *Phys. Rev. B* **84,** 125432 (2011).
18. Cao, L. *et al.* Engineering light absorption in semiconductor nanowire devices. *Nat. Mater.* **8,** 643–647 (2009).
19. Brönstrup, G. *et al.* Optical Properties of Individual Silicon Nanowires for Photonic Devices. *ACS Nano* **4,** 7113–7122 (2010).
20. Spinelli, P., Verschuuren, M. & Polman, A. Broadband omnidirectional antireflection coating based on subwavelength surface Mie resonators. *Nat. Commun.* **3,** 692 (2012).
21. Cao, L. *et al.* Semiconductor nanowire optical antenna solar absorbers. *Nano Lett.* **10,** 439–445 (2010).
22. Diedenhofen, S. L., Janssen, O. T., Grzela, G., Bakkers, E. P. M. & Gómez Rivas, J. Strong geometrical dependence of the absorption of light in arrays of semiconductor nanowires. *ACS Nano* **5,** 2316–2323 (2011).
23. Grzela, G., Hourlier, D. & Gómez Rivas, J. Polarization-dependent light extinction in ensembles of polydisperse vertical semiconductor nanowires: A Mie scattering effective medium. *Phys. Rev. B* **86,** 045305 (2012).
24. Shalev, G., Schmitt, S. W., Embrecht, H., Brönstrup, G. & Christiansen, S. H. Enhanced photovoltaics inspired by the fovea centralis. *Sci. Rep.* **5,** 8570 (2015).
25. Garnett, E. C., Brongersma, M. L., Cui, Y. & McGehee, M. D. Nanowire Solar Cells. *Annu. Rev.*

*Mater. Res.* **41,** 269–295 (2011).
26. Schmitt, S. W. *et al.* Nanowire arrays in multicrystalline silicon thin films on glass: a promising material for research and applications in nanotechnology. *Nano Lett.* **12,** 4050–4054 (2012).
27. Kim, S. K. *et al.* Tuning light absorption in core/shell silicon nanowire photovoltaic devices through morphological design. *Nano Lett.* **12,** 4971–4976 (2012).
28. Alaeian, H., Atre, A. C. & Dionne, J. A. Optimized light absorption in Si wire array solar cells. *J. Opt.* **14,** 024006 (2012).
29. Kim, S.-K. *et al.* Design of nanowire optical cavities as efficient photon absorbers. *ACS Nano* **8,** 3707–14 (2014).
30. Shalev, G., Schmitt, S. W., Brönstrup, G. & Christiansen, S. H. Maximizing the ultimate absorption efficiency of vertically-aligned semiconductor nanowire arrays with wires of a low absorption cross-section. *Nano Energy* **12,** 801–809 (2015).
31. Bohren, C. F. & Huffman, D. R. Absorption an Scattering by an Arbitrary Particle in *Absorption and Scattering of Light by Small Particles* 54-81 (Wiley-VCH, 2007).

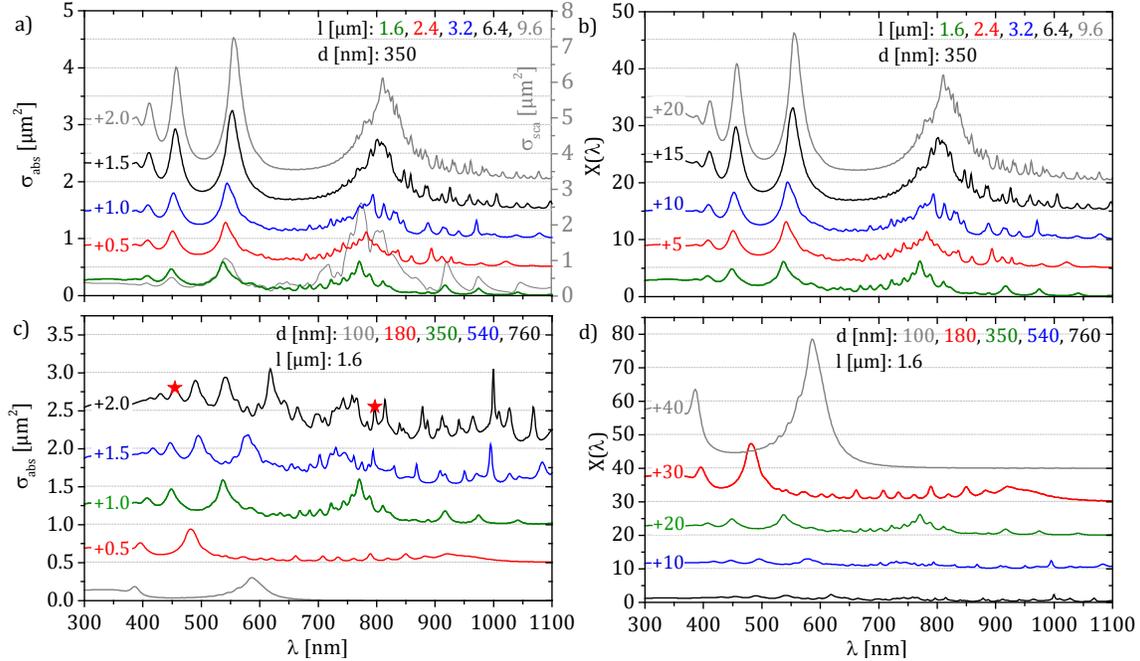

**Figure 1. Spectral absorption cross section and light concentration of individual SiNW. a)** Spectral absorption cross section $\sigma_{abs}$ and **b)** light concentration factor $X$ for free floating individual SiNWs with diameter 350nm and lengths of 1.6, 2.4, 3.2, 6.4 and 9.6µm for a linearly polarized plane wave (300-1100nm) incident along the SiNW axis. The lower grey line in panel a shows the spectral scattering cross section $\sigma_{sca}$ of a SiNW with a diameter of 350nm and a length of 1.6µm (right axis). **c)** Numerically simulated absorption cross section $\sigma_{abs}$ and **d)** light concentration factor $X$ for free floating individual SiNWs with a length of 1.6µm and diameters $d$ of 100, 180, 350, 540 and 760nm for the same irradiation conditions as in panels a and b. The red stars in panel c refer to the inset in Figure 2.

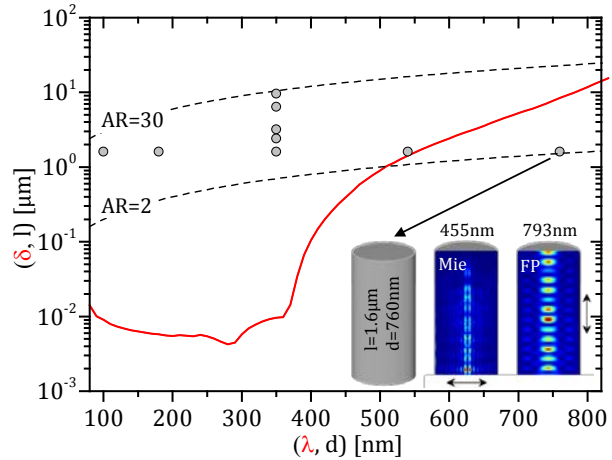

**Figure 2. Spectral absorption depth of light in Si and considered SiNW geometries.** Absorption depth in Si plotted with respect to the wavelength of incident light (red solid line). The same graph illustrates SiNW length over diameter. The grey points indicate the geometries of the SiNWs whose relative absorption enhancement and spectral light concentration is topic of Figure 1. The black dashed lines mark the aspect ratios $AR$ of 2 and 30. The inset depicts the mode formation in a SiNW with $d = 760$nm and $l = 1.6$µm for an incident plane wave with selected wavelengths of 455nm and 793nm along the SiNW axis. The black arrows show the propagation direction of the modes.

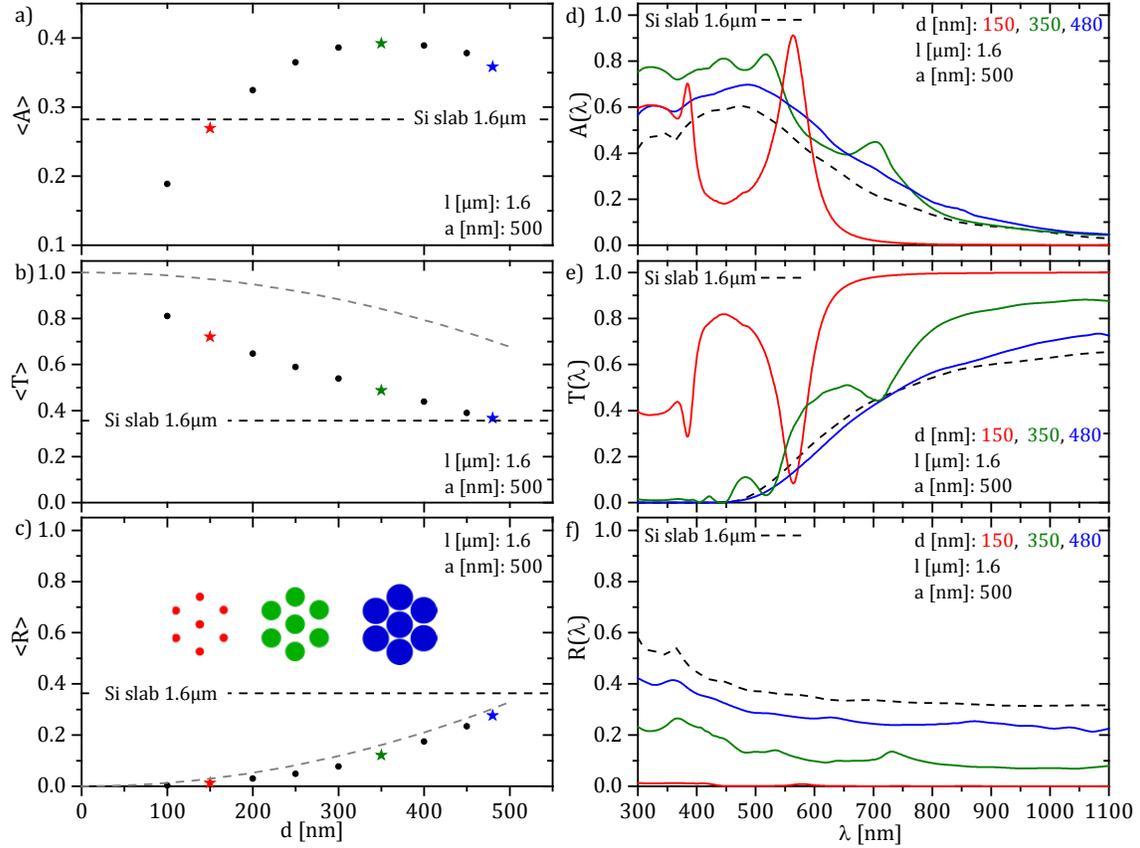

**Figure 3. Light absorption, transmission and reflection of SiNW arrays as function of SiNW diameter. a)** Spectrally averaged light absorption $\langle A \rangle$ **b)** light transmission $\langle T \rangle$ and **c)** light reflection $\langle R \rangle$ of hexagonally aligned SiNW arrays with 1.6μm length and 500nm pitch as a function of SiNW diameter $d$. The black dashed line indicates $\langle A \rangle$, $\langle T \rangle$ and $\langle R \rangle$ of a Si slab with 1.6μm thickness. The grey dashed lines show a linear estimate of $\langle T \rangle$ and $\langle R \rangle$ of the SiNW arrays using the values for the Si slab multiplied by the filling fraction $f$ of the SiNW array with 500nm pitch and the associated diameter. The inset in panel c depicts the top view of array geometries associated with the stars in panels a, b, and c. **d)** Spectral light absorption $A$ **e)** light transmission $T$ and **f)** light reflection $R$ of hexagonally aligned SiNW arrays with 1.6μm length, 500nm pitch and $d$ of 150, 350, and 480nm respectively. The spectra correspond to the stars in panels a, b, and c. The black dashed lines indicate $A$, $T$ and $R$ of a Si slab with 1.6μm thickness.

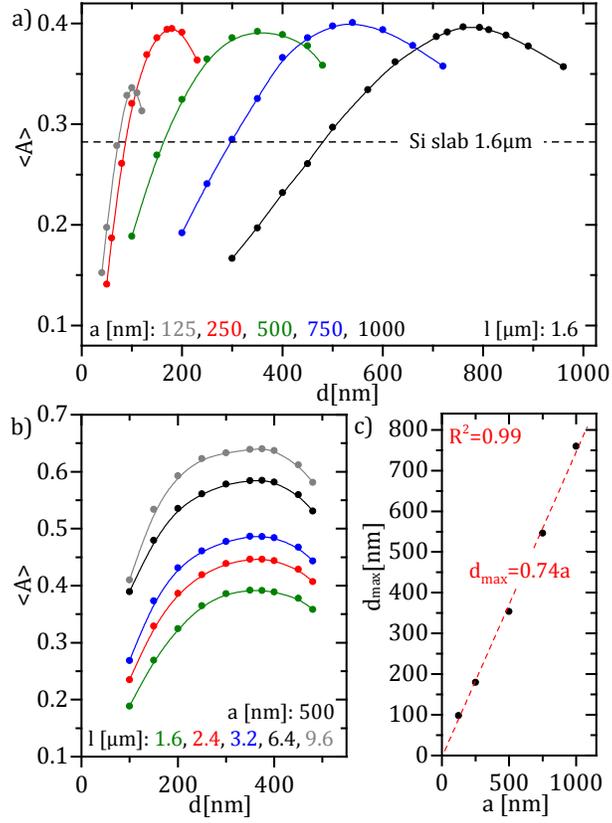

**Figure 4. Maximum spectrally averaged light absorption in SiNW arrays as function of SiNW diameter. a)** Spectrally averaged light absorption $\langle A \rangle$ of hexagonally aligned SiNW arrays with 1.6µm length and a pitch of 125nm, 250nm, 500nm, 750nm, and 1000nm as a function of SiNW diameter $d$. The black dashed line indicates $\langle A \rangle$ of a Si slab with 1.6µm thickness. **b)** $\langle A \rangle$ of hexagonally arranged SiNW arrays with a pitch of 500nm and a length $l$ of 1.6µm, 2.4µm, 3.2µm, 6.4µm, and 9.6µm respectively. **c)** SiNW diameter at the maximum $\langle A \rangle$ of a SiNW array with length 1.6µm ($d_{max}$) as a function of pitch $a$. The dashed red line indicates a linear fit passing through the origin.

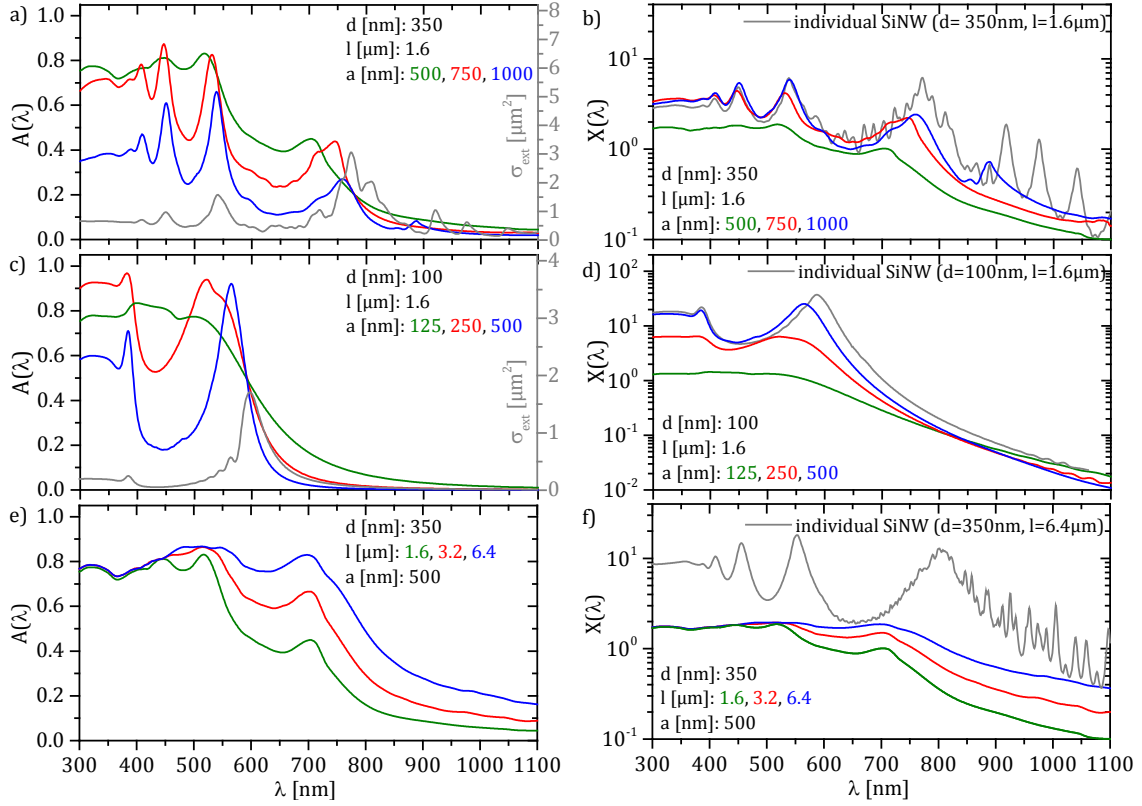

**Figure 5. Spectral light absorption and concentration in SiNW arrays as function of array pitch. a)** Spectral light absorption $A$ and **b)** spectral light concentration $X$ of hexagonally aligned SiNW arrays with 1.6µm length, 350nm diameter and a pitch of 500, 750 and 1000nm, respectively. The gray line in panel a shows the spectral extinction cross section $\sigma_{sca}$ of an individual SiNW with 1.6µm length and a diameter of 350nm. The gray line in panel b shows the spectral spectral light concentration $X$ of the same individual SiNW. **c)** Spectral light absorption $A$ and **d)** spectral light concentration $X$ of hexagonally aligned SiNW arrays with 1.6µm length, 100nm diameter and a pitch of 125, 250 and 500nm, respectively. The gray line in panel c shows the spectral absorption cross section $\sigma_{sca}$ of an individual SiNW with 1.6µm length and a diameter of 100nm. The gray line in panel d shows the spectral spectral light concentration $X$ of the same individual SiNW. **e)** Spectral light absorption $A$ and **f)** spectral light concentration $X$ of hexagonally aligned SiNW arrays with 500nm pitch, 350nm diameter and a length of 1.6, 3.2 and 6.4µm, respectively. The gray line in panel f shows the spectral spectral light concentration $X$ of an individual SiNW with 6.4µm length and a diameter of 350nm.

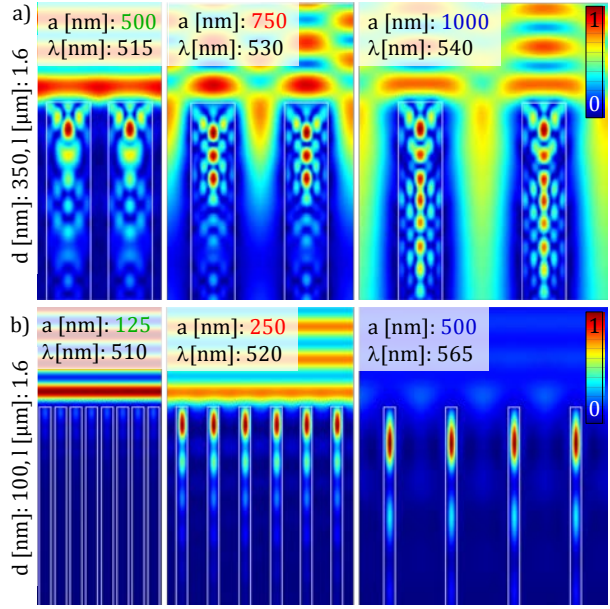

**Figure 6. Dependence of photonic mode formation in SiNW arrays on the array pitch. a)** Relative cross sectional energy density ($E^2$) of adjacent SiNW in a hexagonally aligned SiNW array with 1.6μm length, 350nm diameter $d$ and a pitch $a$ of 500, 750, and 1000nm. Images are recorded at a wavelength of 515, 530, and 540 nm, respectively, associated to peaks in the related absorption and concentration spectra in Figures 5a and b. **b)** Relative cross sectional energy ($E^2$) of adjacent SiNWs in a hexagonally aligned SiNW array with 1.6μm length, 100nm diameter $d$ and a pitch $a$ of 125, 250, and 500nm. Images are recorded at a wavelength of 510, 520, and 565 nm, respectively, associated to peaks in the related absorption and concentration spectra in Figures 5c and d.

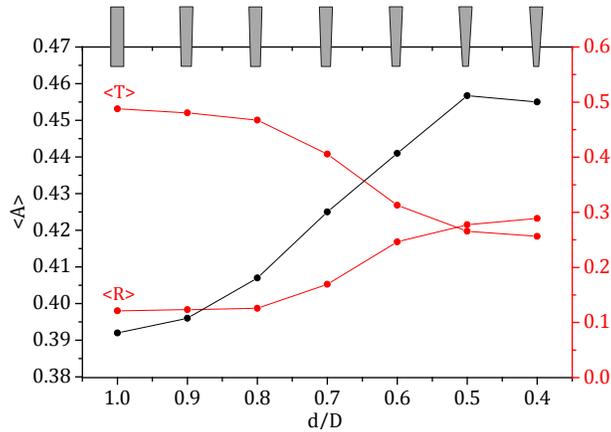

**Figure 7. Spectrally averaged light absorption in SiNW / SiNC arrays as function of the SiNW / SiNC bottom diameter.** Average light absorption $\langle A \rangle$, transmission $\langle T \rangle$ and reflection $\langle R \rangle$ of SiNWs / SiNCs with 350nm diameter and 1.6μm length arranged in a hexagonal lattice with a pitch of 500nm as a function of SiNWs / SiNCs bottom diameter. The bottom diameter $d$ is plotted in units of the top diameter $D$. Insets on top show the schematic cross sections of the SiNWs / SiNCs with the associated geometries.

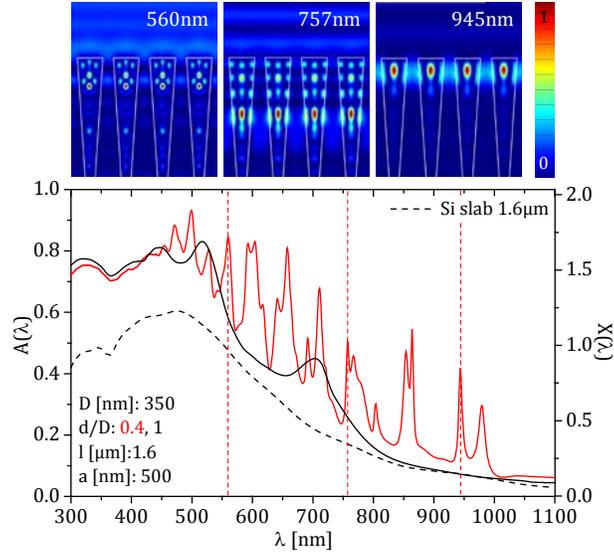

**Figure 8. Spectral light absorption of SiNW as compared to SiNC arrays.** Spectral light absorption $A$ (concentration $X$) of SiNWs with 350nm diameter and 1.6µm length arranged in a hexagonal lattice with a pitch of 1000nm (black) and for the same array with a reduced SiNW bottom diameter $d/D = 0.4$ (red). The dashed black line shows $A$ of a 1.6µm thick Si slab for reference. Images on top represent the color-coded relative cross sectional energy density in three adjacent SiNCs of the array with reduced bottom diameter for the absorption maxima at 560nm, 757nm, and 945nm marked by the red dashed lines.

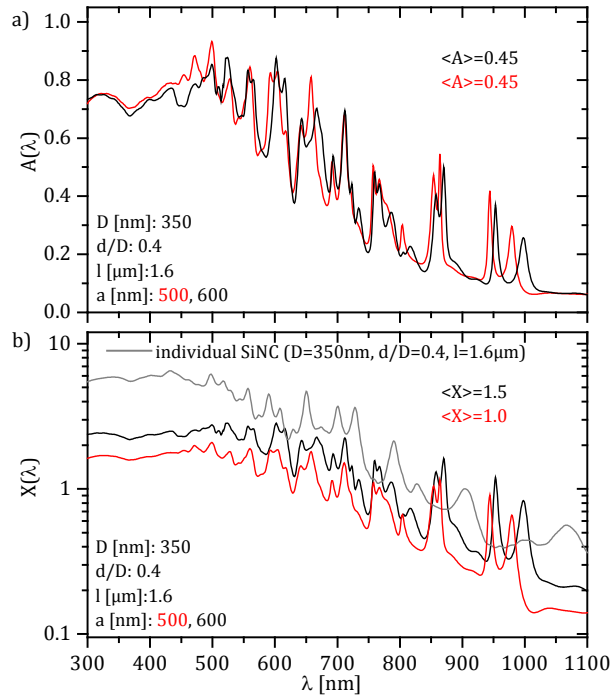

**Figure 9. Spectral light absorption and concentration in SiNC arrays as function of array pitch. a)** Spectral light absorption $A$ of SiNCs with 350nm top diameter, bottom diameter $d/D = 0.4$ and 1.6µm length arranged in a hexagonal lattice with a pitch of 500nm (red) and 600nm (black). **b)** Spectral light concentration $X$ of SiNCs with 350nm upper diameter, bottom diameter $d/D = 0.4$ and 1.6µm length arranged in a hexagonal lattice with a pitch of 500nm (red) and 600nm (black). The grey line shows $X$ of an individual SiNC with 350nm top diameter, bottom diameter $d/D = 0.4$ and 1.6µm length.

17